# Physical conditions in the ISM towards HD185418

Gargi Shaw[1], G. J. Ferland[1], R. Srianand[2], and N. P. Abel[1]

## Abstract

We have developed a complete model of the hydrogen molecule as part of the spectral simulation code Cloudy. Our goal is to apply this to spectra of high-redshift star-forming regions where $H_2$ absorption is seen, but where few other details are known, to understand its implication for star formation. The microphysics of $H_2$ is intricate, and it is important to validate these numerical simulations in better-understood environments. This paper studies a well-defined line-of-sight through the Galactic interstellar medium (ISM) as a test of the microphysics and methods we use. We present a self-consistent calculation of the observed absorption-line spectrum to derive the physical conditions in the ISM towards HD185418, a line-of-sight with many observables. We deduce density, temperature, local radiation field, cosmic ray ionization rate, chemical composition and compare these conclusions with conditions deduced from analytical calculations. We find a higher density, similar abundances, and require a cosmic ray flux enhanced over the Galactic background value, consistent with enhancements predicted by MHD simulations.

Subject headings: ISM: abundances, ISM: clouds, ISM: structure, ISM: individual (HD185418).

[1] University of Kentucky, Department of Physics and Astronomy, Lexington, KY 40506; gshaw@pa.uky.edu, gary@pa.uky.edu, npabel2@uky.edu

[2] IUCAA, Post bag 4, Ganeshkhind, Pune 411007, India; anand@iucaa.ernet.in

# 1. Introduction

Understanding the physical conditions in the interstellar medium (ISM) and the sources that maintain these conditions are very important for understanding galaxies and their evolution. Most of our present day understanding of the ISM of our Galaxy is based on physical quantities derived from the absorption lines seen in the spectra of bright stars. By analogy, absorption lines seen in spectra of high-redshift quasars can reveal the conditions in young star-forming galaxies at intermediate redshift, where few other observables are present (Wolfe et al. 2003; Srianand et al. 2005a).

Rotational excitations of $H_2$, fine-structure excitations of species such as C I, O I, Si II and C II, and relative populations of elements in different ionization states are used to infer the kinetic temperature (Savage et al. 1977), the UV radiation field (Jura 1975), gas pressure (Jenkins & Tripp 2001), particle density, and the ionization rate in the ISM of our Galaxy. Most of the attempts to derive the physical quantities are based on simple analytical prescriptions. Considerable insight can be gained by interpreting the observations using a self-consistent calculation that takes into account all the physical processes (see, for instance, van Dishoeck & Black 1987; Gry et al. 2002).

Damped Ly$\alpha$ systems seen in the spectra of QSOs are believed to originate from high redshift galaxies. A minority of these absorbers, about 15% of the "damped Ly$\alpha$ absorbers", show $H_2$ and C I absorption lines (Petitjean et al. 2000; Ledoux et al. 2003; Srianand et al. 2005a). The availability of good spectroscopic observations covering a wide wavelength range allows one to create a detailed model of these systems. We have included a detailed microphysical simulation of $H_2$ (Shaw et al. 2005 and the references cited there) into the spectral simulation code Cloudy (Ferland et al. 1998). The goal is to use $H_2$ to understand processes and conditions in these intermediate redshift galaxies (Shaw et al. 2005; Srianand et al. 2005b). One aim of this paper is to validate our calculations in a nearby known environment and use this code at high redshift where very little is known.

Here we present a self-consistent calculation of the thermal, ionization, and chemical state and the resulting spectrum, with the aim of deriving the physical conditions in the Galactic ISM towards the star HD185418. This has a very well characterized line-of-sight with many observables (Sonnentrucker et al. 2003, hereafter S03). Here we interpret the observed spectrum and compare conclusions from the numerical simulations with other known properties of the sight line.

This paper is organized as follows. We first summarize the observed data along the line-of-sight HD185418 in Section 2 and describe other boundary conditions that influence our calculations in Section 3. We first compute properties of a cloud with the temperature, column density, and composition deduced by S03, but with the density suggested from C I excitation. This calculation fails to produce the column densities of C II*, H I, and high rotational levels of $H_2$. Next the constant temperature assumption was relaxed and thermal equilibrium calculations presented in Section 3.4. This produced a temperature a factor of two lower than observed. We next vary the hydrogen density, ionization radiation, abundances and cosmic ray flux to reproduce the observed values. A cosmic ray ionization rate 20 times higher than the Galactic background is required. This calculation reproduces most of the observed column densities. This is followed by a demonstration of the influences of these free parameters on the observed spectrum, to identify the observational consequences of each physical process in Section 3.5. We conclude with a discussion and summary of conclusions in Section 4.

## 2. A well-characterized line-of-sight

HD185418 is a well studied B0.5 V star located at Galactic coordinates $(l, b) = (53°, -2.2°)$ at a distance of 790 pc from the sun. This line-of-sight has a large number of molecular, atomic, and ionic absorption lines.

S03 gather together extensive observational data and derive column densities from spectra with the Far Ultraviolet Spectroscopic Explorer (FUSE) and Hubble Space Telescope/Space Telescope Imaging Spectrograph (HST/STIS). Table 1 summarizes these column densities.

The measured $E(B-V)$, 0.50 (Fitzpatrick & Massa 1988, 1990; S03) can be converted into a total hydrogen column density of $N(H) \approx 2.9 \times 10^{21}$ cm$^{-2}$ for an assumed dust-to-gas ratio of $A_v/N(H) = 5.30 \times 10^{-22}$ (Draine 2003). This is roughly consistent with the measured column densities of $H^0$ (from Ly$\alpha$) and $H_2$. Here we define the hydrogen molecular fraction as $f(H_2) = 2N(H_2)/[N(H^0)+ 2N(H_2)]$. The observed log $[N(H_2)]$ and log $[N(H^0)]$ are 20.71±0.15 and 21.11±0.15 respectively, and H is nearly half molecular ($f(H_2) = 0.44$).

This line-of-sight shows various Lyman and Werner band absorption lines of $H_2$ (summarized in Table 1). $J = 5$ is the highest detected rotational level and lines from excited vibrational levels are not detected. Observers usually derive an excitation temperature from the ratio of column densities of $J = 1$ and $J = 0$, defined as

$$N(J=1) = 9N(J=0)\exp(-170.5/T_{10}) \quad [\text{cm}^{-2}]$$

or,

$$T_{10} = -170.5\left\{\ln\left[\frac{N(J=1)}{9N(J=0)}\right]\right\}^{-1} \quad [\text{K}]$$

(1.)

This is often assumed to be the average kinetic temperature (Savage et al. 1977). S03 find $T_{10}$ to be about 100 ± 15K. Rachford et al. (2002) found a mean temperature of 68 ± 15 K for Galactic lines of sight with $N(H_2) > 10^{20.4}$ cm$^{-2}$. In their sample most of the sight lines have $T_{10} < 75$ K and only 3 out of 23 sightlines (including HD 185418) have $T_{10} > 95$ K. In the sample of Savage et al. (1977) also we notice that the mean $T_{10}$ is 55 ± 8 K for sight lines with $N(H_2) > 10^{20.4}$ cm$^{-2}$. None of these 8 sight lines have $T_{10}$ greater than 80 K. Clearly HD 185418 seems to be one of the very few systems that show high $T_{10}$ with high $H_2$ column density.

Molecules detected along this line-of-sight include CO, CN, $CH^+$, CH, and $C_2$. The column densities of these species define the chemical state of the gas. Various atomic and ionic lines arising from C, O, S, Si, Mg, K etc. are also seen. Some, for instance C I, include excited fine-structure column densities, making it possible to derive the local pressure and hydrogen density ($n_H$) (Jenkins & Tripp 2001; Srianand et al. 2005a). S03 obtain the mean $n_H \sim 6.3 \pm 2.5$ cm$^{-3}$ for an assumed kinetic temperature of 100 K. A simple analytical calculation of C II and C II* yields a local electron density ($n_e$) equal to 0.002 cm$^{-3}$. However, $n_e$ derived assuming photoionization equilibrium between the neutral and singly ionized species is much higher ($0.03 < n_e$ (cm$^{-3}$) $< 0.37$) if the radiation field is the Galactic background (Draine 1978).

S03's fits to the absorption line profiles of K I, S I, C I*, CI**, O I*, CO and CH suggests that 3 main components of molecular gas with a velocity spread of 4.5 km s$^{-1}$ exist along the line-of-sight. The absorption lines of other species spread over a velocity range of 15 km s$^{-1}$ in 9 distinct components. However, we notice that most of the column densities of Na I and Ca II reside in the three main components noted above. S03 find fractional abundances of carbon in three components that are, within observational uncertainties, identical. This means that the physical conditions are roughly identical in these three components. The absorption lines of CH and $CH^+$ are detected in two of these components. The velocity dispersion within each component is small, and the Doppler $b$ parameters are $\sim$ 0.5 to 1.6 km s$^{-1}$. In addition, the presence of S III, Si III, N II absorption lines suggests that $\sim$ 1% of the gas along the line-of-sight is ionized.

Our goal in the remaining sections is to reproduce these observed column densities, derive physical conditions using the methods applied at high redshift, and compare these conclusions with known properties of this line-of-sight.

## 3. Calculations

This section describes various calculations and compares the predicted column densities with the observed ones. All the calculations are done with the spectra simulation code Cloudy (05.08). The code was last described by Ferland et al. (1998), Abel et al. (2004), and Shaw et al. (2005; hereafter S05).

Our calculation is based on energy conservation and chemical balance. The temperature is derived from heating and cooling balance, including various processes such as gas and grain photo-electric heating, cosmic ray heating, heating due to $H_2$ dissociation and collisional de-excitation, and cooling via fine-structure atomic and molecular lines. Ionization and electron density are determined from balancing ionization and recombination processes.

Our calculations are non-equilibrium, but assume that atomic processes are time-steady. That is, the density of a species or level is given by a balance equation of the form

$$\frac{\partial n_i}{\partial t} = \sum_{j \neq i} n_j R_{ji} + Source - n_i \left( \sum_{j \neq i} R_{ij} + Sink \right) = 0 \ [\text{cm}^{-3} \ \text{s}^{-1}].$$

Here $R_{ji}$ represents the rate [s$^{-1}$] that a species $j$ goes to $i$, *Source* is the rate per unit volume [cm$^{-3}$ s$^{-1}$] that new atoms appear in $i$, and *Sink* is the rate [s$^{-1}$] they are lost. This, together with equations representing conservation of energy and charge, fully prescribes the problem.

We use the $H_2$ chemistry network, consisting of various state-specific formation and destruction processes, described by S05. In a cold and dusty medium $H_2$ is formed mainly on grain surfaces, whereas in a hot dust-free medium it is formed via associative detachment ($H^- + H^0 \rightarrow H_2 + e^-$). These exothermic formation processes produce $H_2$ in excited vibrational and rotational levels, often referred to as formation pumping. $H_2$ is destroyed mainly via the Solomon process when the gas is optically thin to the $H_2$ electronic lines. Most excitations of the excited electronic states of $H_2$ result in decays to the highly excited vibrational and rotational levels of the ground electronic state (Solomon pumping), which further decay down via quadruple transitions. We also consider the vibrational and rotational excitations of $H_2$ by cosmic rays.

Our detailed chemical network is discussed in Abel et al. 2005. We have included photoionization and ionization by cosmic rays, collisional ionization, Compton scattering of bound electrons, and Auger multi-electron ejection. We have also included radiative recombination, low-temperature dielectric recombination, and charge exchange reactions with both gas and dust. The column densities of singly ionized species in our calculations are decided by various processes listed above and not only by the ionization – radiative recombination equilibrium.

Our grain physics includes three chemical species, each resolved into a number of size bins. It determines the grain charges and photoelectric heating self-consistently. Details about the grain physics of our code are given by van Hoof et al. (2004). Grain temperatures are combined with the temperature-dependent formation rates of Cazaux & Tielens (2002) to derive total $H_2$ formation rates. Our heavy-element chemistry network consists of nearly 1050 reactions with 71 species involving hydrogen, helium, carbon, nitrogen, oxygen, silicon, sulphur, and chlorine (Abel et al. 2005).

### *3.1. Cloud geometry*

We consider a plane parallel geometry with radiation striking from both sides. Earlier van Dishoeck and Black (1986) had used a similar geometry to study physical conditions of the diffuse interstellar clouds. Our calculations, which follow this pioneering work, find photo-interaction rates by carrying out explicit integrations of atomic and molecular cross sections over the local radiation field. This field includes the attenuated incident continuum and the diffuse emission from all gas and grain constituents. Because of this the full incident continuum, from very low to very high energies, must be explicitly specified. The Galactic background radiation field given by Black (1987) is the only source of photoelectric heating and ionization. This radiation field includes the Cosmic Microwave Background (CMB) at $T = 2.7$ K, background radiation in the infrared, visible, and ultraviolet as tabulated by Mathis, Mezger, & Panagia (1983); and the soft X-ray background described by Bregman & Harrington (1985). We parameterize the intensity of the incident radiation by $\chi$, the ratio of the assumed incident radiation field to the Galactic background.

Our interpolated continuum extends across the full spectral region. Actually photoelectric absorption by the ISM removes photons in the energy range 1 to roughly 4 Ryd. We reproduce this effect by extinguishing the radiation field by a cold neutral column density with column density $N_{ext}(H) = 10^{22}$ cm$^{-2}$. Tests show that the exact value of $N_{ext}(H)$ has little effect on column densities within various rotational levels of $H_2$ for a given total $H_2$ column density. This is because the levels are predominantly pumped by Balmer continuum photons with $h\nu < 1$ Ryd. However, the ionization potentials of $N^0$ and $S^+$ are 1.068 and 1.715 Ryd respectively so the populations of these species are sensitive to this continuum.

The aim of this paper is to test the theoretical tools we have developed for high-redshift DLAs and apply them to neutral and molecular sight lines through our Galaxy. This sight line also has N II, Si III and S III absorption lines that suggest the presence of a warm ionized medium. The Galactic background radiation field we use is assumed to have had most of the H-ionizing radiation field removed and so is incapable of producing these ions. The ionization potentials suggest that these ions will exist within an H II region, which may be physically associated with the background B0.5 star. Tests show that a 29,000 K B0 star is capable of producing significant amounts of these ions. In the following we will list our predicted S III and N II column densities but will not attempt to reproduce the observed values.

We assume constant density as would be the case when magnetic or turbulent pressure dominates the gas equation of state. The gas ionization and temperature depend on the shape and intensity of the incident radiation field, the gas density, and total column density. Thus, in this type of model, a single cloud will have many different regions, an ionized and warm neutral region near the surface, a cold neutral medium at deeper depths, with a largely molecular core in shielded regions.

Cosmic rays play a crucial role in various interstellar processes, producing heating in ionized gas, ionization in neutral gas, and driving ion-molecule chemistry. Our treatment of cosmic-ray heating is described in Abel et al. (2005). We assume an $H^0$ ionization rate ($\Gamma_{CR}$) of $2.5\times10^{-17}$ s$^{-1}$ (Williams et al. 1998) as the Galactic background value in atomic regions. Enhanced cosmic ray densities can occur near regions of active star formation since the rate is a balance between new cosmic rays produced by supernovae and their loss through several processes. In the following calculations we will vary the cosmic ray ionization rate to match our predicted column densities with the observed ones. We represent the cosmic ray ionization rate as $\chi_{CR}$, the rate relative to $\Gamma_{CR}$. For reference, McCall et al. (2003) find $\chi_{CR} = 48$ for one line of sight through the nearby ISM towards $\zeta$ Persei. Liszt (2003) also finds enhanced cosmic-ray ionization rate compared to the Galactic background.

Our calculations, like all those that use a standard ISM chemistry network in simulations of the diffuse ISM, under predict abundances of CH and $CH^+$. This has been noticed by others groups (Kirby et al. 1980; Gredel 1997). It is known (Draine & Katz 1986; Zsargó & Federman 2003) that non-equilibrium chemistry can be an important channel for production of CH and $CH^+$. Alternatively, these species can be produced in shocked gas, or in a warmer gas phase (Gry et al. 2002). Furthermore, the rate coefficients may be in error. In the following we will list our predicted $CH^+$ and CH column densities but will not be able to reproduce the observed values. We plan to test our chemical network for $CH^+$ and CH in future paper.

The S03 data show three main velocity components along the line-of-sight. However, our calculations use a single layer of gas with stratified regions. This is justified because of the near constancy of C I*/C I found in the three main components (S03), which suggests that their thermal pressures are comparable. The physical state of the gas is mainly determined by the total gas column density and grain optical depths, and the resulting continuum extinction and line shielding. Furthermore, we use the observed total column densities since column densities for each individual component are not available.

## *3.2. Microturbulence*

Microturbulence plays an important role in determining the optical depth of a line. Increased microturbulence decreases the line-center optical depth of $H_2$ lines and increases the line width. As a result more continuum photons are absorbed and the Solomon pumping rate increases. The observed $b$ value for the $H_2$ lines is $6.2 \pm 0.5$ km s$^{-1}$. We adopt 6 km s$^{-1}$ microturbulence in all our calculations. This is not actually a physical microturbulence but rather includes macroscopic motions of the clouds, but is necessary to correctly account for the line self-shielding.

There are two types of self-shielding, continuum and line. Small velocity shifts have no effect on the continuum self-shielding and do not depend on column densities of individual clouds. Line self-shielding depends on velocity shifts of individual clouds. However, if we use the right total line width this accounts for the presence of clouds at slightly different velocities. Cloudy calculates line self-shielding in a very accurate self-consistent way. So, our single cloud approximation will not affect the conclusions. However, we can model multi-component environments if we have detailed information on each component.

Constant density is assumed, as would be the case when magnetic or turbulent pressure dominates the gas equation of state. The strength of the magnetic field is not known along this line-of-sight, but the ratio of magnetic to gas pressure is generally large in the ISM in those cases where it is known (Heiles & Crutcher 2005). The supersonic *b* values quoted above would correspond to a turbulent pressure greater than the gas pressure. In these cases a constant density model also has constant pressure (Tielens & Hollenbach 1985).

## *3.3. Constant-temperature calculations*

As a first test case we assume the density (6 cm$^{-3}$) and temperature (100 K) derived by S03. We consider a Galactic background radiation field with $\chi = 1$, a Galactic cosmic ray ionization rate of $\chi_{CR} = 1$, and the ISM gas-phase abundances of Savage & Sembach (1996). The chemical, ionization, and level population equilibrium are determined using this assumed temperature. This is a simplification since the temperature does vary across a cloud in most physical situations. We stop our calculation at $N(H) = 10^{21.46}$ cm$^{-2}$, as determined from the extinction and a typical extinction per hydrogen column density. The main purpose of this exercise is to perform calculations around the best fitted values obtained by S03.

This calculation under predicts the column densities of excited fine-structure levels of C I, C II, O I and Si II, all of which are density sensitive. The predicted column densities are given in column three of Table 1. The model calculations over predict the carbon ionization and under predict the C I fine-structure excitations. We analytically recomputed the density from the C I fine-structure excitations and find $10 < n_H$ (cm$^{-3}$) $< 20$ for $T = 100$ K. We included collisions by H$^0$ and H$_2$, UV pumping with the rate given by Silva & Viegas (2000), and pumping due to the CMB. We will take $n_H = 15$ cm$^{-3}$ as a first guess at the density.

The predicted column densities assuming the higher density and ISM abundances (see the footnote of table 2) but with other parameters held at the S03 values are given in column four of Table 1. These predictions are consistent with most of the observed values (S03), suggesting that the density and temperature are an appropriate starting point. Some significant discrepancies exist. We find the correct $N(C\ I^*/C\ I)$, $N(C\ I^{**}/C\ I)$ and $T_{01}$ as expected. However, it over predicts the column densities of lower rotational levels of H$_2$, under-predicts H$_2$ in $J > 2$ and the H I column density. The predicted $N(C\ II^*)/N(C\ II)$ is higher than the observed value. The predicted column densities of C I, Mg II, Fe II, Ni II and Mn II are slightly higher than the observed values. This suggests that the depletion of these elements along the line-of-sight is higher than the ISM values we assumed. This is consistent with Joseph et al.'s (1986) suggestion that depletions are greater along more reddened lines-of-sight.

The next step is to reproduce the temperature of the cloud using self-consistent thermal equilibrium calculations. This is important for understanding the cloud's heating sources which in turn will influence the populations of excited states, the main observational diagnostic of the gas.

## *3.4. Thermal equilibrium calculations*

Here we balance heating and cooling to determine the temperature.

First we try to match the observed column densities by varying the hydrogen density, the radiation parameter $\chi$, and the abundances of the heavy elements. All calculations had a total $N(H) = 2.9 \times 10^{21}$ cm$^{-2}$ and $\chi_{CR} = 1$. These calculations produced a kinetic temperature of typically $T \sim 50$ K, substantially below the value of 100 K deduced from $T_{10}$, and also failed to produce enough excited state populations in many ions and in $J > 1$ rotational levels of H$_2$. However, the constant-temperature calculations discussed in the previous section show that these excited state column densities will be reproduced if additional heating processes can raise the gas temperature to ~100 K.

This sight line is warmer than expected for clouds with similar $N(H_2)$. The measured $T_{10}$ is more than 2$\sigma$ higher than the mean (68±15 K) measured along lines-of-sight with similar extinction (Rachford et al. 2002). This suggests that some additional heating sources may exist.

We tried raising the background radiation field. This did raise the temperature but it also increased the ionization of the gas, conflicting with the observations. We also know from IRAS observations that HD 185418 does not interact significantly with the absorbing gas (S03). It is most likely that the UV field is not much higher than the Galactic background.

We also did tests which included polycyclic aromatic hydrocarbons (PAHs), which are known to be the dominant source of photoelectric heating in some clouds. We assume an empirical PAH density law, $n(PAH) \propto n(H^0)/n(H)$, as described in Abel et al. (2005). This is suggested by observations showing that PAHs are a molecular cloud surface phenomenon, destroyed in the H$^+$ region (Giard et al. 1994) and coagulated into larger grains in fully molecular gas (Jura 1980). PAHs have little effect on the H$_2$ temperature because of their assumed low abundance in molecular gas. We also computed an extreme case with a constant density of PAHs with ISM abundances, but found, as expected, that they have a profound effect on the chemistry of molecular regions. We do grain heating, temperature, and charging fully self-consistently (van Hoof et al. 2004). When PAHs are abundant in molecular gas they remove nearly all free electrons, which strongly change the chemical balance. None of these tests succeeded in raising the temperature of the H$_2$ region by significant amounts. This is not considered further.

Next we treat the cosmic ray ionization rate as a free parameter. We find that $\chi_{CR} = 20$, a gas density of $n_H = 27$ cm$^{-3}$, and $\chi = 1.1$ produces the observed temperature, column densities in rotational levels of H$_2$, and other atom and ionic fine structure excitations and column densities. These results are presented in column five of Table 1.

In Figure 1 we plot the column densities for various $J$ levels as a function of the excitation temperature. The filled circles and triangles represent the predicted and the observed values. The open circles represent local thermodynamic equilibrium (LTE) column densities obtained by assuming that the level populations are given by Boltzmann statistics for the temperature at each point in the cloud. It is clear that the $J = 0-1$ levels are in LTE and hence $T_{10}$ can be safely used to determine the weighted mean temperature. Our predicted $T_{10}$ is 74 K while the observed $H_2$ temperature is 100±15 K, and the predicted $H^0$-weighted temperature is 79 K. Grain photoionization and cosmic ray interactions are the main sources of heating. The cooling is dominated by the [C II] 158 μm and [O I] 63 μm lines. Hydrogen is the dominant electron donor even in deep regions of the cloud, due to efficient cosmic ray ionization.

Although we have assumed cosmic rays, any source of heat that contributes $2.7\times10^{-25}$ erg cm$^{-3}$ s$^{-1}$ without altering the ionization of the gas would produce many of the same effects. Actually this is an old problem in the ISM literature – ISM gas cooling rates deduced from the [C II] 158 μm line has long been known to exceed the known heating sources (Pottasch et al. 1979). But, as noted above, the sight line is unusually warm for its measured extinction, so it must have some unusual property.

Our calculations reproduce the observed column densities of all the high $J$ levels within the observational uncertainties. The derived value of $n_H$ is well below the critical density required for thermalizing the $J > 2$ levels of $H_2$. Higher rotational levels are populated by Solomon pumping and cosmic ray excitation and as a result the column density in the $J = 4$ and 5 levels are greater than their LTE values.

S03 derived electron densities from both the ionization and excitation of the gas. They found that a wide range of electron densities, between 0.002 and 0.32 cm$^{-3}$, were required to reproduce these observed value. Our calculations predict the radial dependence of the radiation field, density, ionization, chemistry, and temperature (Figures 2a, b and 3). Our model simultaneously reproduces the excited fine-structure level populations of C I, C II, O I, the observed column density ratios of trace elements, and the ionization ratios of $N$(C I)/$N$(C II), $N$(S I)/$N$(S II), and $N$(Fe I)/$N$(Fe II). We consider collisional excitations by H, $H_2$, He, $H^+$, and $e^-$ for fine structure excitations of O I and collisional excitations by H (Barinovs et al. 2005), and $e^-$ (Dufton et al. 1994, At. Data Nucl. Data Tables) for fine structure excitations of Si II. [Si II] λ 34.5 μm can originate in either the H II region or in the neutral PDR as the ionization potential of Si is less than 13.6 eV. Figures 2a, 2b, and 3 show the ionization structures and temperature profile for this best-fit case. The structure is very similar to a classical PDR (Tielens & Hollenbach 1985). The predicted electron density ranges from 0.047 to 0.026 cm$^{-3}$ and the temperature ranges from 102 to 65 K across the cloud.

The best-fit abundances are summarized in Table 2. The abundances of C, O, Ca, Fe and Mg are similar to those of S03. Our predicted O I, C I*, and C I** column densities match well with the observed data. In the absence of good constraints on the trace elements, S03 have assumed ISM abundances for S and K. Our models require depletion factors of 0.5 for S from ISM abundance.

Our calculations reproduce the observed column densities of Cl I and Cl II. The dominant $Cl^0$ recombination process is charge exchange of $H_2$ with $Cl^+$, which forms $HCl^+$, which in turn produces $Cl^0$. In the absence of this channel $Cl^0$ would be under predicted in our models. We have included ~ 30 reactions involving chlorine and its ions with rates taken from UMIST database. Our detailed study of $Cl^0$ and $Cl^+$ will be given in a separate paper.

We reproduce the observed column density of CO. Initially, we used the UMIST chemical reaction network and our predicted CO column density was 0.5 dex smaller than the observed value. There have been many improvements in the recent years in the study of photodissociation of CO (Federman et al. 2001, 2003) and it is known that the experimental oscillator strengths are larger than those used in UMIST. This creates a faster dissociation rate when the gas is optically thin to the CO electronic lines, and more self-shielding when the lines become optically thick. However, self-shielding is only important above column densities of about $N(CO) \sim 10^{15}$ cm$^{-2}$. Experimental oscillator strengths are not available for the majority of the CO electronic lines, establishing an uncertainty in the predictions. The improved reaction rate given by Dubernet et al. (1992) in an important CO production channel ($C^+ + OH \rightarrow CO + H^+$) increases the CO column density by $\approx$ 0.5 dex. The column density of CO along this line-of-sight is consistent with these new rates.

The predicted column densities of CN and $C_2$ are consistent with the observed upper limits. However, our calculations under-predict the column densities of CH and $CH^+$. As mentioned previously, this is a general problem in calculations of interstellar chemistry that has been noted by other groups. We use the UMIST rate for the reaction $CH + H \rightarrow C + H_2$. This is an important destruction channel for CH. We also tried with a rate $2.7 \times 10^{-11}(T/300)^{0.38}\exp(-2200/T)$ (private communication with E. Roueff). The new rate with temperature barrier did increase the CH column density but still it was less than the observed value.

We also predict the column densities of Ne I, Ne II, Si I, Mg I, OH, $H_3^+$ and HCl, although these have not yet been observed along this line-of-sight. The predicted column densities and their abundances are listed in Table 1 and 2. The predicted $H_3^+$ column density offers a way to test our conclusion that the cosmic ray ionization rate is high along this sight line.

Figures 4a shows the transmitted continuum in the wavelength range 0.09-0.13 μm computed for our best fit model. There are thousands of $H_2$ electronic lines in this range which are strongly overlapped. Such simulated spectra make it possible to take unknown line blends into account. Figures 4b shows this transmitted continuum in higher resolution in the wavelength range 0.105-0.110 μm.

## *3.5. Variation of parameters around the best value*

This section shows how the observed column densities change with variations in the parameters around the best-fitting values. We vary each of the parameters ($n_H$, ionizing radiation, cosmic ray ionization rate, the UV radiation field, and the turbulence) while keeping all other parameters fixed to show the physical consequences.

Figure 5 shows the effects of varying $n_H$. The $H_2$ formation rate depends on $n_H$, so more $H_2$ is produced as $n_H$ is increased. A higher $H_2$ column density produces more self-shielding and results in an over population of the $J=0$ level. The higher fine-structure levels of C I are collisionally pumped by $H^0$, $H_2$ and $e^-$. In a neutral medium the collisional excitation is dominated by $H^0$ and $H_2$ collisions. As a result, C I*/C I and C I**/C I increases with increasing $n_H$. It is clear that $n_H$ in the range 13-27 cm$^{-3}$ can explain the observed ratios of $N$(C I*)/$N$(C I), $N$(C I**)/$N$(C I), $H_2(J=0)/H_2(J=1)$ and $H_2(J=0)/H_2(J=3)$. We also notice that $N$(O I*)/$N$(O I) can be reproduced in this range of $n_H$. The range is higher than that found from analytical estimates by S03. However, it is consistent with our analytical estimate given in section 3.3. Thus, in this system the C I fine-structure level populations, combined with $T_{10}$, gives the correct gas density.

We found that cosmic rays play an important role in heating and ionizing the gas. Figure 6 shows the effects of a range of cosmic ray ionization rates on the column density ratios of $N$(C I*)/ $N$(C I), $N$(C I**)/ $N$(C I) and the rotational levels of $H_2$. Cosmic rays increase the densities of $H^0$, $H^+$, which undergo exchange collisions with $H_2$ and induce ortho-para conversions. Cosmic ray ionization rates 10-20 times the Galactic background value given by Williams et al. (1998), but half that found by McCall et al. (2003) in one sight line, are required to reproduce the temperature and fine-structure level populations of C I. An enhancement greater than 10 is required to reproduce the observed $J=1/J=0$ (or $T_{10}$) ratio of $H_2$. As discussed above, the mean kinetic temperature of the gas in this range is very close to $T_{10}$, as expected in a molecular region that is shielded from the Solomon process.

Figure 7 shows the effects of changing the Galactic background radiation, i.e. $\chi$, on column density ratios. The observed range in $f(H_2)$ constrains $1 < \chi < 2.5$. Most of the singly ionized species have only limits to the column densities and so they do not provide additional constraints on the radiation field. We notice that the value of $\chi$ can not be much larger than 1 as the predicted column density ratios of C I*/C I and C I**/C I increase for larger $\chi$.

The rate of formation of $H_2$ on dust has the greatest uncertainty in our calculations. The Jura rate ($3\times10^{-17}$ cm$^3$s$^{-1}$) is generally taken as the standard, although there are significant variations (Browning et al. 2003). Figure 8 shows the $H_2$ density as a function of depth for two different rates. As expected, the cloud is more molecular for larger Jura rate. A factor of 2.7 change in the rate changes $N(H_2)$ by a factor of 1.96. Figure 9 shows the effects of varying the formation rate on the column densities of various rotational levels of $H_2$. An increased formation rate causes $H_2$ to form at smaller column densities where the Solomon process is faster. As a result, the total $H_2$ column density and the level populations in various rotational levels, which are excited by the Solomon process, increase with an increased formation rate. However, formation rates in the range of 2.7 - $4.2\times10^{-17}$ cm$^3$s$^{-1}$ reproduce the observed H$_2$ column densities for the other rotational levels.

## 4. Discussion & Conclusions

We have reproduced most of the column densities observed by S03 along the line-of-sight towards HD185418. This suggests that our detailed treatment of the microphysics does capture the critical physics of such clouds in the ISM and validates the application to higher redshift objects. Other conclusions include the following:

- Our best-fit model is one which follows the changes in the temperature and ionization as a function of cloud depth. The best-fit parameters are $n_H = 27$ cm$^{-3}$, $\chi_{CR} = 20$, and $\chi = 1.1$. The H$^0$-weighted average temperature is $\sim 79$ K while $T_{10}$ is 74 K. The electron density ranges from 0.047 to 0.026 cm$^{-3}$.

- The calculation predicts the variation in physical conditions across the cloud and offers insights into basic heat sources. The temperature, electron fraction, chemical state, and local radiation field is determined self-consistently. While our overall results ($T$, $n_H$, and metal abundance) are in general agreement with those found by analytical theory, our detailed simulations do find some differences. Many of these differences are caused by the need for the analytical calculation to make ad hoc assumptions about the relative densities of colliders that produce line excitations, or in the variation of these collider densities and their temperature over the cloud.

- A UV radiation field and cosmic ray ionization rate similar to the mean values seen in diffuse interstellar medium do not reproduce the observed $T_{10}$. Comparison with other sight lines shows that this sight line is unusually warm for gas with its extinction and $N$(H$_2$). Clearly an additional heat source is needed to explain the observed $T_{10}$.

- We find that increasing the UV radiation field cannot provide the required additional heating since it also increases the ionization. A very large overabundance of PAHs would be needed for this to account for the extra heating. We consider this to be ad hoc and do not pursue it.

- Cosmic rays can provide additional heat without drastically altering the chemical and ionization balance of the cloud. We find that a CR ionization rate 20 times the rate obtained by William's et al. (1998) is required. They derived this rate for very dense molecular clouds with $n$(H$_2$) in the range of $1\times10^4$ - $3\times10^4$ cm$^{-3}$. McCall et al. (2003) derived an cosmic ray flux towards $\zeta$ Persei enhanced by 48 times this, based on their laboratory study of H$_3^+$- e$^-$ recombination rate. Our derived rate is intermediate. In general the cosmic ray ionization rate will depend on the production rate, propagation loses, confinement time-scale, and configuration and strengths of magnetic fields. Padoan & Scalo (2005) have shown that self-generated MHD waves produce an enhanced cosmic-ray density in diffuse clouds compared to those found in dense clouds. This suggests that enhanced rates might be found in lower density clouds like the sight line studied here.

- Our conclusion that more heat is needed is robust, although our suggestion of enhanced CRs is only one possibility. Our calculations establish that an extra heating rate of $2.7\times10^{-25}$ erg cm$^{-3}$ s$^{-1}$ is needed to account for the observed temperature. Gry et al. (2002) argue that shocks can produce heating and enhance the column densities of high $J$ levels of $H_2$ and $CH^+$. Other possibilities include a strongly enhanced abundance of PAHs or other sources of mechanical or photo energy.

- We also present predicted column densities for species that have not yet been observed. The predicted $N(H_3^+)$ offers a test of our conclusion that cosmic rays are enhanced (McCall et al. 2003).

- A comparison between theoretical and observed column densities of species with only one ion stage observed permits the abundances of those species to be derived.

## 5. Appendix

Here we mention the uncertainties in the atomic data which affect our predicted column densities of Fe and K.

The rate coefficient for $Fe^0 + H^+ \rightarrow Fe^+ + H^0$ is uncertain by at least a factor of 2. We find three values in the published literatures. Pequignot & Aldrovandi (1986) give a rate coefficient $3.\times10^{-9}$ cm$^3$ s$^{-1}$ whereas UMIST and Tielens & Hollenbach use $7.4\times10^{-9}$ cm$^3$ s$^{-1}$. We use an intermediate value $5.4\times10^{-9}$ cm$^3$ s$^{-1}$ as listed in ORNL webpage (http://cfadc.phy.ornl.gov/astro/ps/data/home.html). Test shows that UMIST rate gives better agreement with the observed column density of $Fe^0$. As a result, the column density of Fe I as listed in table 1 of this paper has an uncertainty of at least by a factor of 1.6.

In this model, $K^0$ is a trace stage of ionization due to its low ionization potential. Most of potassium is $K^+$. The photoionization cross-section for $K^0$ is taken from Verner et al. (1995). Verner et al. (1995) give analytical fit to photoionization cross sections calculated by the Hartree-Dirac-Slater (HDS) method. This method can be inaccurate for neutrals and first ions near the threshold of outer shells (Verner et al. 1996) and this will propagate into the prediction of K I column density listed in table 1 of this paper. Verner et al. (1996) fit the photoionization cross section of atoms and ions of elements (H, He, N, O, Ne, C, Na, Mg, Al, Si, S, Ar, Ca, Fe) done by the Opacity Project (Seaton et al. 1992). They have plotted the differences in photoionization cross sections of C, Si, and Ar due to these two different calculations. The Opacity Project (OP) calculates the photoionization cross sections based on the $R$-matrix method and hence gives more accurate low-energy photoionization cross sections.

We compare the photoionization cross sections of the elements calculated by Verner et al. in 1996 (based on OP) and 1995 (based on HDS). We find two distinct categories. (i) In the first group, the ratios of photoionization cross sections for the valence shells vary smoothly over a small range 0.1 to 10. Elements like, $C^0$, $O^0$, $N^0$, $S^0$, and $Al^0$ fall in this group. These have 2p and 3p configurations for their valence electrons. (ii)In the second group, there is a large minimum in the cross section, there is a shift in energy between the two calculations, and the value of the minimum photoionization cross section differs by orders of magnitude. As a result, the ratios of photoionization cross sections are very large at the position of minimum photoionization cross section. $Na^0$ and $Mg^0$ show this distinct feature. The valence electron configuration of $Na^0$ and $Mg^0$ are $3s$ and $3s^2$. We expect the same features to be present in $K^0$ which has valence electron configuration of $4s$. However, we do not know this for sure since experiment and accurate theory have not been done.

Keeping these large uncertainties in mind we consider rescaling the photoionization cross sections of the outer shell for the atoms not considered by OP. We rescale the valence shell photoionzation cross section of $K^0$ by 5 to match the observed K I column density. This rescaling method is a guess and we recommend that low-energy photoionization cross sections should be calculated by more accurate $R$-matrix method.

**Acknowledgements**

We thank John Black and S. R. Federman for valuable discussions on CO chemistry. We also thank T. P. Snow and D. Welty for valuable suggestions. This work is supported by the Center for Computational Sciences at the University of Kentucky, the NSF through grants AST 03 07720 and INT-0243091, NASA through grant NAG5-12020, and STScI through AR 10653 and HST-AR-10316. GJF and RS acknowledge the support from DST/INT/US (NSF-RP0-115)/2002.

We would like to thank the anonymous referee for his/her thoughtful suggestions.

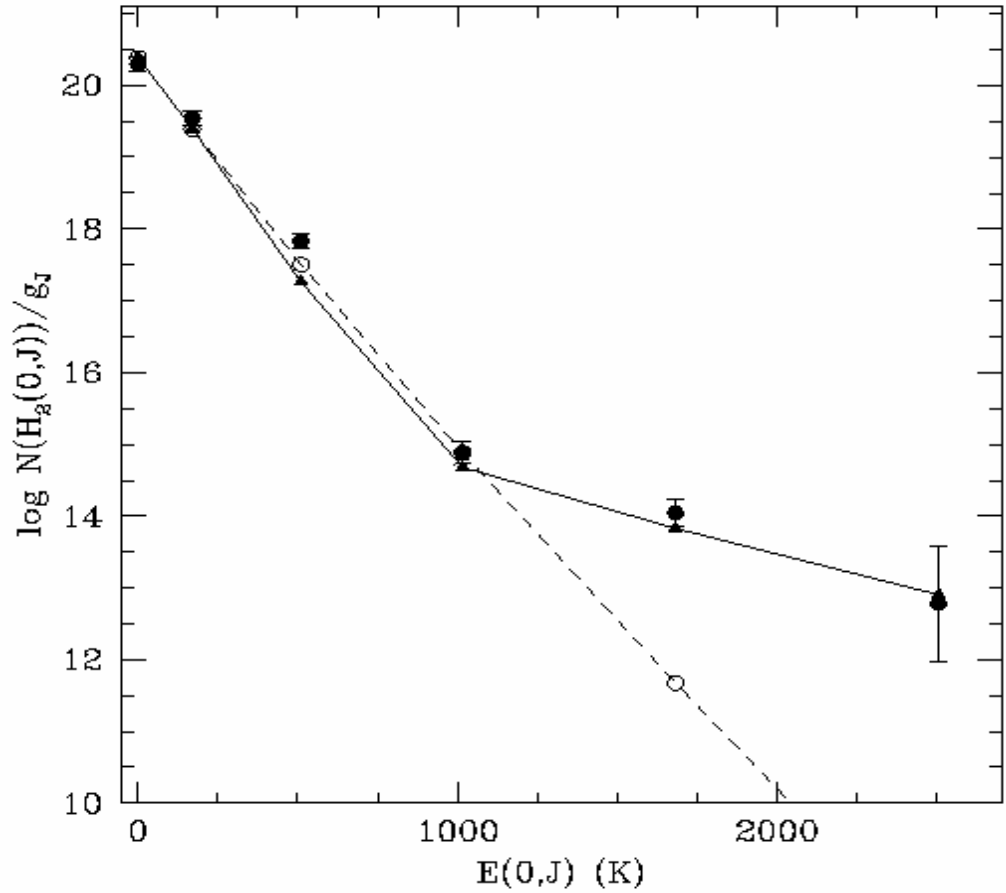

FIGURE 1. $H_2$ excitation diagram. The $H_2$ column densities divided by corresponding statistical weights for different rotational levels are plotted as a function of excitation temperature. The filled circles, open circles and triangles are from observations (S03), obtained under LTE assumption and our best-fit model respectively.

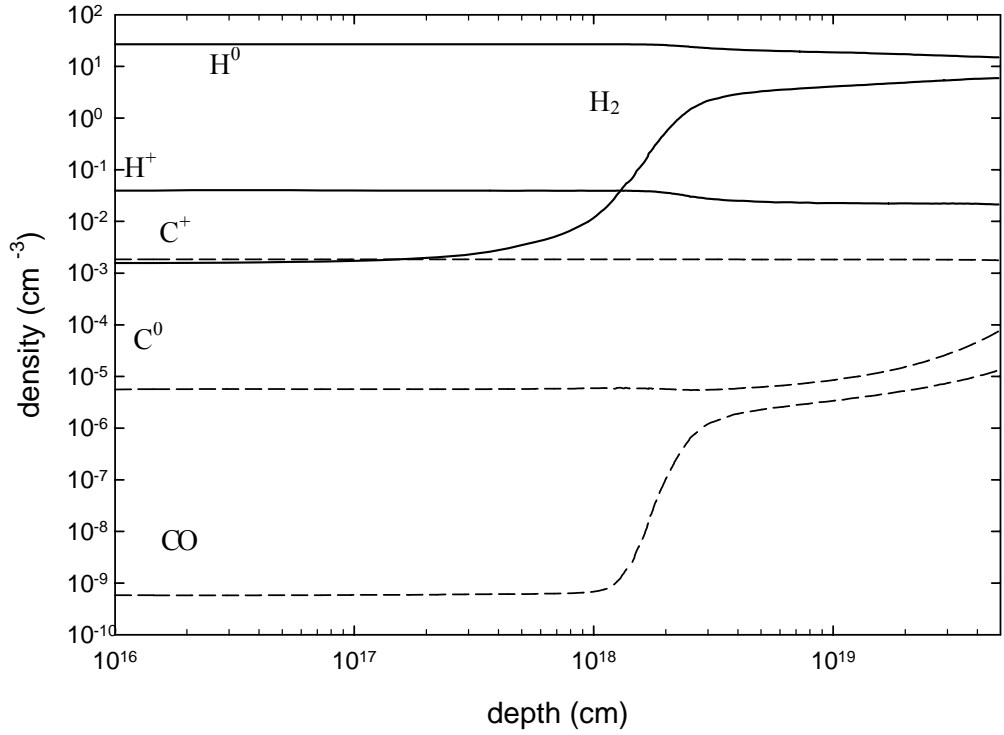

FIGURE 2a: The hydrogen and carbon ionization and chemical structure for the best-fit model along the line-of-sight HD185418. This plot represents one half of the cloud. The other half of the cloud is symmetrical to this half.

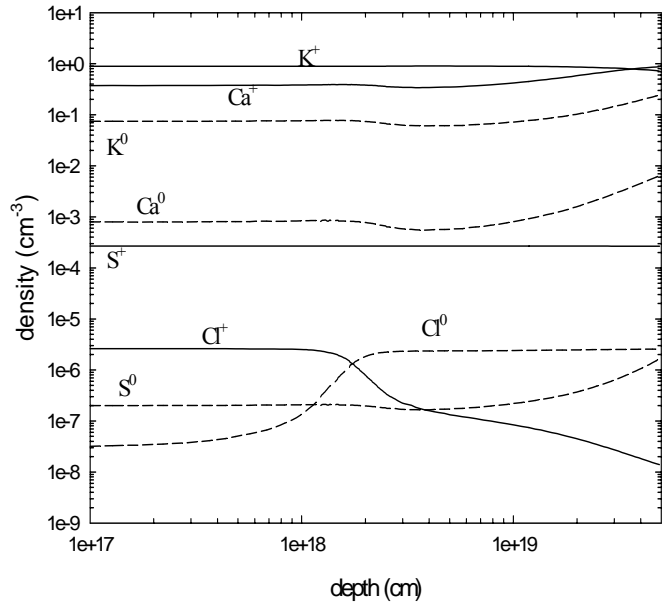

FIGURE 2b: The ionization structure of various neutral and singly ionized species for our best-fit model. This plot represents one half of the cloud. The other half of the cloud is symmetrical to this half.

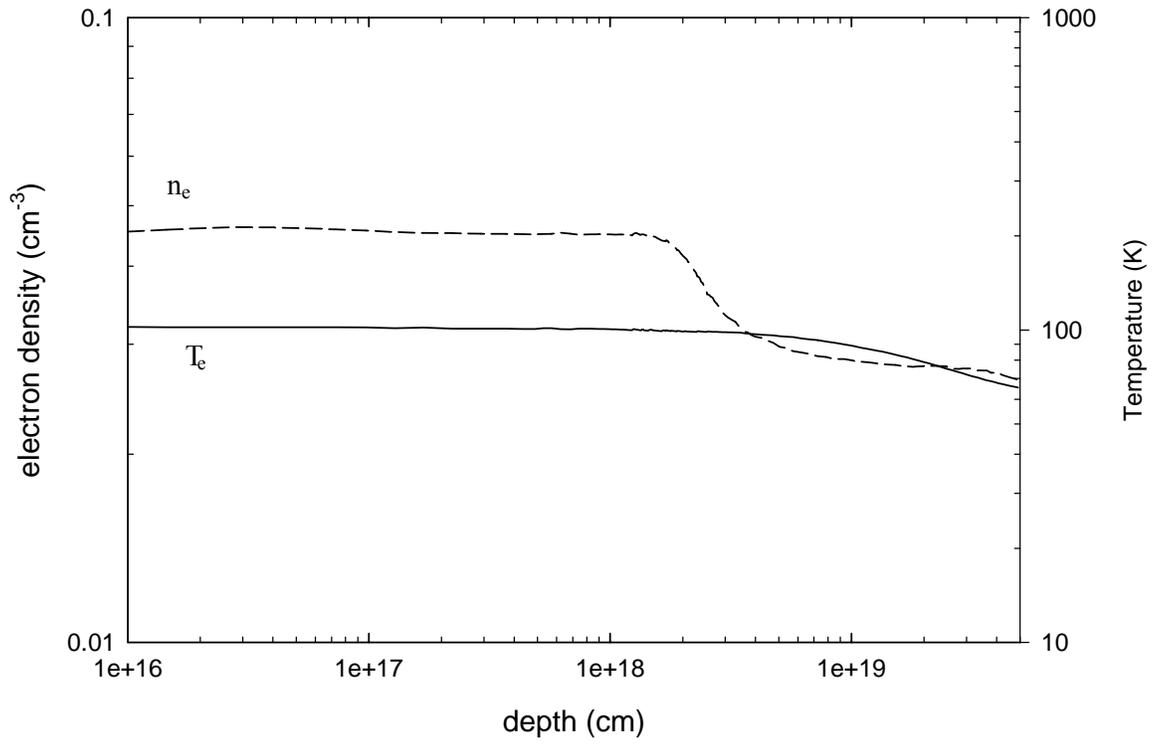

FIGURE 3. The electron density and temperature structure for our best-fit model. This plot represents one half of the cloud. The other half of the cloud is symmetrical to this half.

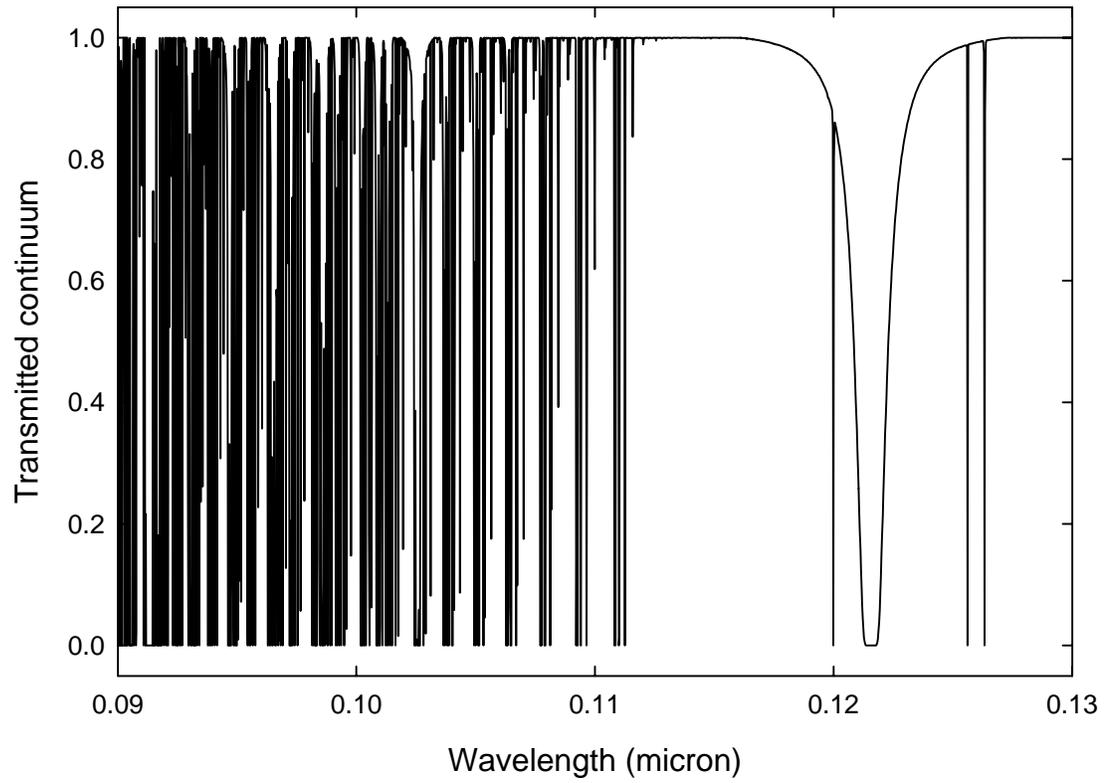

FIGURE 4a. The transmitted spectrum for our best fitted model is shown.

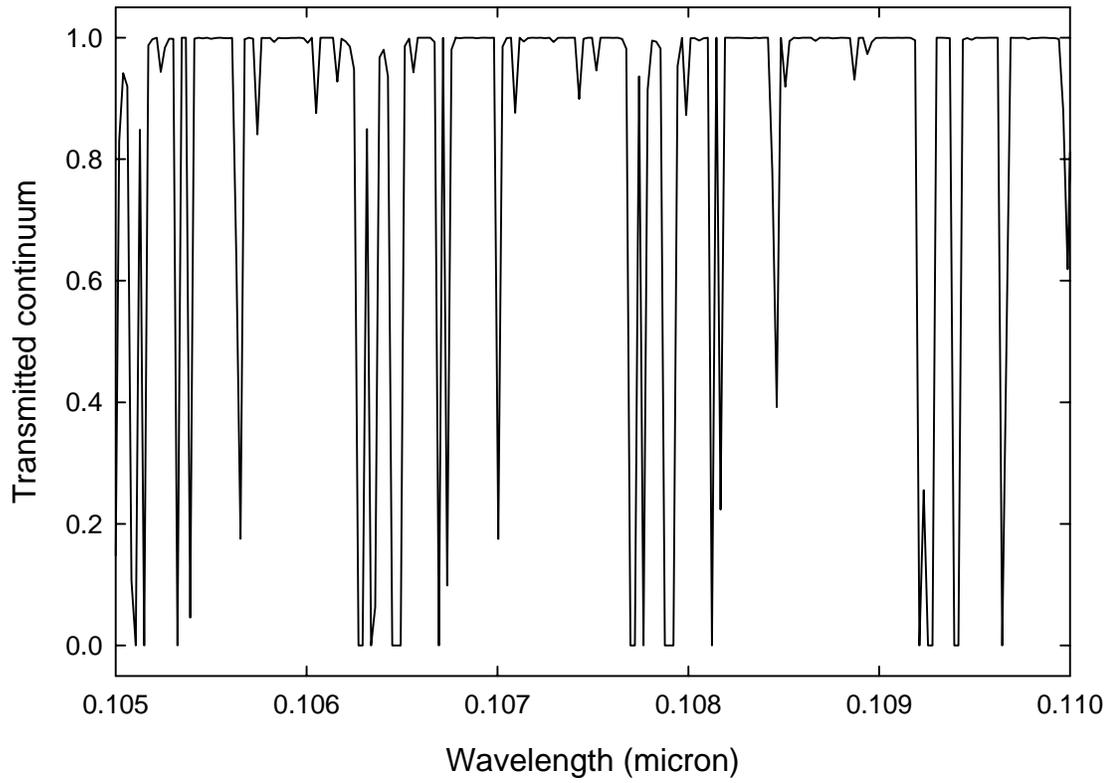

FIGURE 4b. The transmitted spectrum for our best-fit model is shown. The wave length range plotted most of the absorption line seen are Lyman band $H_2$ lines from the ground vibrational states.

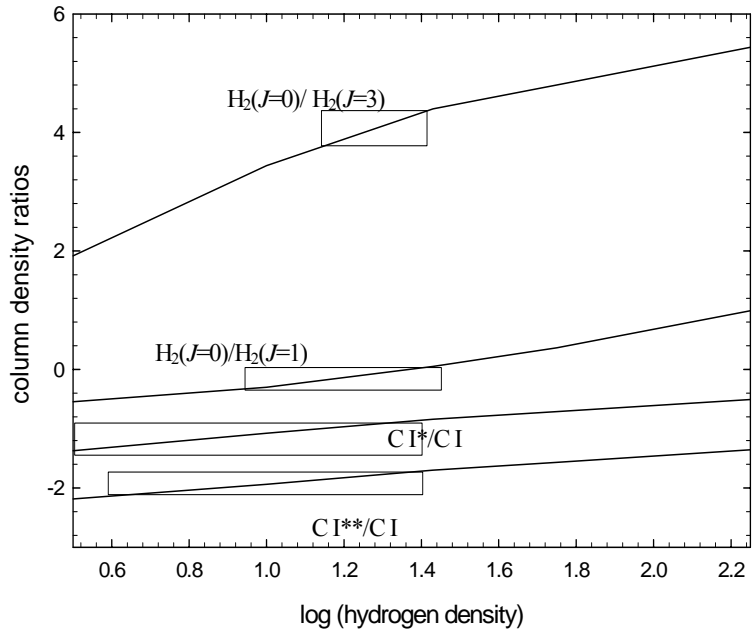

FIGURE 5. The effects of varying the hydrogen density around the best-fit value are shown. The curves represent the model predictions and the box in each curve represents the range in $n_H$ that produces the observed ratios within the measurement uncertainties quoted in S03.

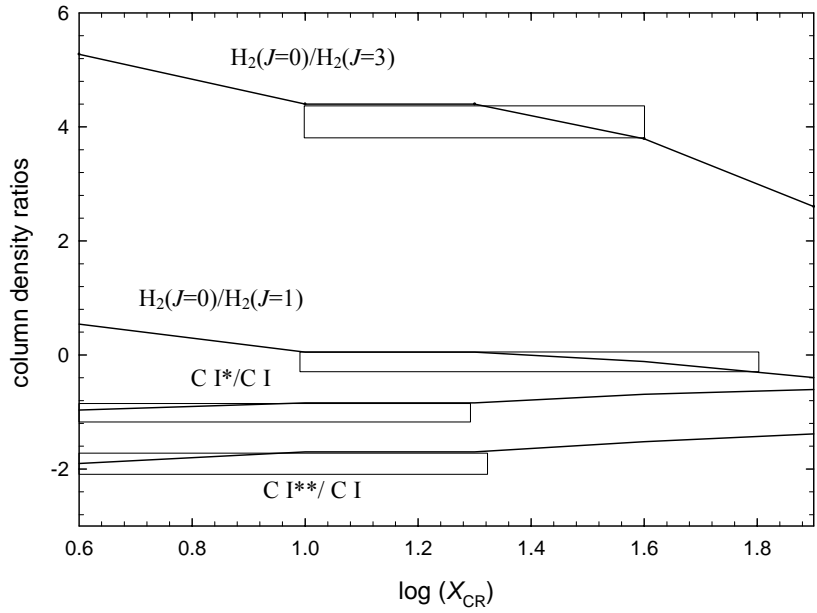

FIGURE 6. The effects of varying the cosmic ray ionization rate around the best-fit values are shown. The curves represent the model predictions and the box in each curve represents the range in $n_H$ that produces the observed ratios within the measurement uncertainties quoted in S03.

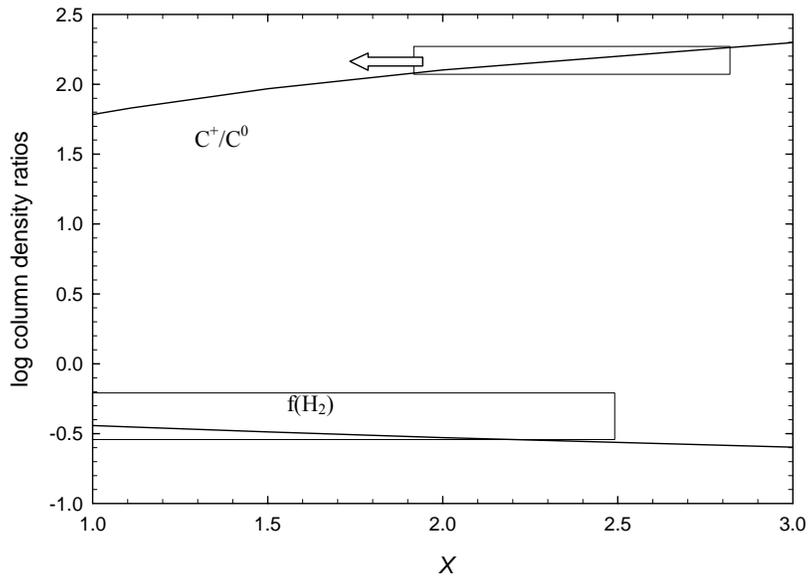

FIGURE 7. The effects of varying the UV radiation field around the best fitted value are shown. The solid curves represent our predictions and the box in each curve represents the range in cosmic ray ionization rate which is consistent with the observational uncertainties. In the case of $C^+/C^0$ only upper limit is available.

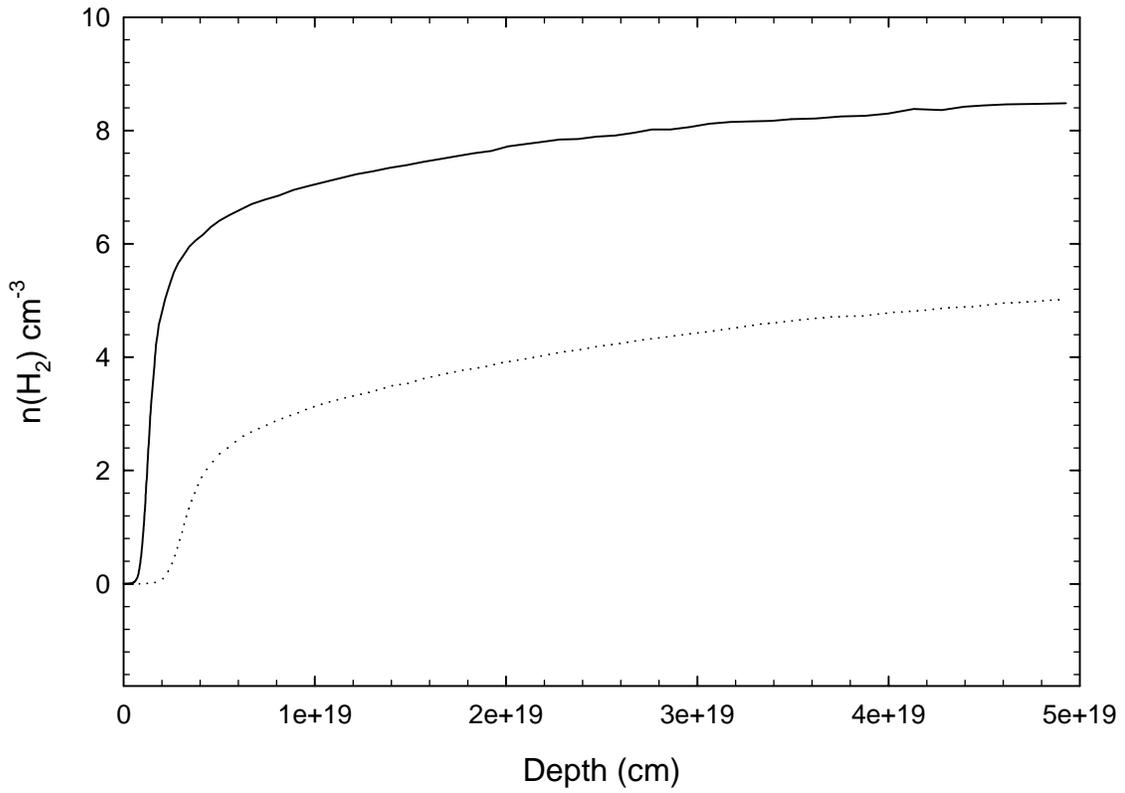

FIGURE 8. $H_2$ density is shown as a function of depth for 0.75 times (dotted line) and two times (solid line) Jura rate along the line-of-sight HD185418. This plot represents one half of the cloud. The other half of the cloud is symmetrical to this half.

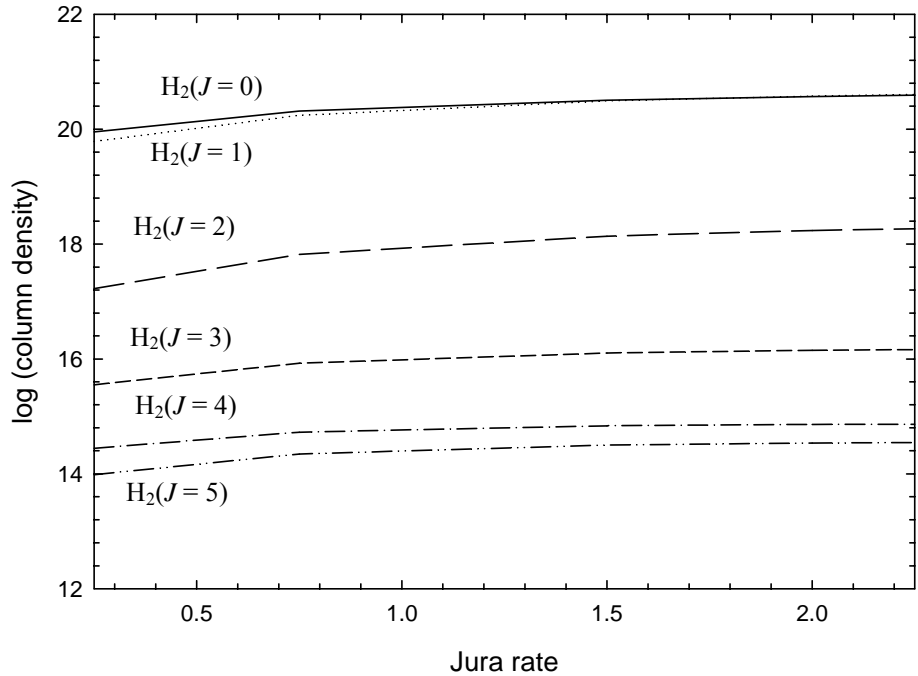

FIGURE 9. The effects of varying the scaling of the Jura-rate on $H_2$ column density along the line-of-sight HD185418 are shown. The solid ($J = 0$), dotted ($J = 1$), long-dashed ($J = 2$), short-dashed ($J = 3$), dash-dot ($J = 4$) and dash-dot-dot ($J = 5$) lines represent different rotational levels of $H_2$.

Tables:

Table 1

The column densities for observed and different investigated models along the line-of-sight towards HD185418

| Chemical species | log $N$(cm$^{-2}$) | | | |
|---|---|---|---|---|
| | Observed | Constant $T$ $n_H$ = 6 cm$^{-3}$ | Constant $T$ $n_H$ = 15 cm$^{-3}$ | Thermal equilibrium |
| C I | 15.53±0.09 | 14.96 | 15.19 | 15.36 |
| C I* | 14.45±0.08 | 13.46 | 13.97 | 14.52 |
| C I** | 13.59±0.08 | 12.58 | 13.14 | 13.66 |
| O I | 18.15±0.09 | 17.93 | 17.93 | 18.28 |
| O I* | 12.61±0.16 | 11.65 | 12.03 | 12.38 |
| O I** | ≤ 12.40 | 11.29 | 11.77 | 11.79 |
| CO | 14.70±0.10 | 14.24 | 14.40 | 14.82 |
| H$_2$($J$=0) | 20.30±0.10 | 20.43 | 20.59 | 20.40 |
| H$_2$($J$=1) | 20.50±0.10 | 20.61 | 20.78 | 20.35 |
| H$_2$($J$=2) | 18.34±0.10 | 17.93 | 18.38 | 17.96 |
| H$_2$($J$=3) | 16.20±0.15 | 15.44 | 15.73 | 16.00 |
| H$_2$($J$=4) | 15.00±0.20 | 14.05 | 14.17 | 14.78 |
| H$_2$($J$=5) | 14.30±0.80 | 13.65 | 13.76 | 14.42 |
| H I | 21.11±0.15 | 21.12 | 20.84 | 21.24 |
| S I | 13.66±0.08 | 13.30 | 13.53 | 13.80 |
| S II | >15.36 | 16.94 | 16.94 | 16.43 |
| S III | 13.81±0.07 | 13.14 | 12.93 | 12.83 |
| Si II | >14.20 | 15.93 | 15.93 | 15.93 |
| Si II* | 11.72±0.18 | 9.91 | 10.10 | 10.41 |
| N I | 17.30±0.09 | 17.33 | 17.33 | 17.33 |
| N II | >14.40 | 11.23 | 10.37 | 11.37 |
| Fe II | 14.93±0.10 | 15.23 | 15.23 | 14.83 |
| Fe I | 11.84±0.08 | 12.10 | 12.36 | 12.10 |
| Mg II | 16.02±0.13 | 16.53 | 16.53 | 16.12 |
| Mg I | | 13.91 | 14.14 | 14.47 |
| Ar I | >13.77 | 15.88 | 15.88 | 15.88 |
| C II | ≤ 17.75 | 17.83 | 17.83 | 17.26 |
| CII* | 14.93±0.10 | 15.10 | 15.40 | 15.09 |
| CN | ≤ 11.70 | 9.39 | 10.20 | 10.01 |
| CH | 13.11±0.05 | 12.10 | 11.68 | 11.48 |
| $CH^+$ | 13.12±0.09 | 10.51 | 10.68 | 9.62 |
| $C_2$ | ≤ 13.02 | 5.95 | 7.16 | 5.86 |
| Cu II | 12.49±0.07 | 12.60 | 12.60 | 12.60 |
| Ni II | 13.50±0.07 | 13.69 | 13.69 | 13.68 |
| Mn II | 13.61±0.10 | 13.79 | 13.69 | 13.79 |
| Ca I | 10.30±0.05 | 7.90 | 8.33 | 10.26 |
| Ca II | 12.62±0.05 | 11.30 | 11.47 | 12.64 |
| K I | 11.88±0.03 | 10.98 | 11.21 | 11.93 |
| Cl I | 14.52±0.16 | 14.29 | 14.33 | 14.38 |
| Cl II | < 13.40 | 13.64 | 13.24 | 13.15 |
| Ne I[†] | | | | 17.52 |
| Ne II[†] | | | | 14.86 |
| OH | | | | 15.09 |
| $H_3^+$ | | | | 13.19 |
| HCl | | | | 13.03 |

† Assuming ISM abundances

Table 2

Comparison of derived parameters for HD185418

| Parameters | S03 | Constant $n_H$ |
|---|---|---|
| $n_H$(cm$^{-3}$) | 6.3±2.5 | 27 |
| $\chi$ | not specified | 1.1 |
| $T_{10}$ (K) | 100±15 | 74 |
| $\chi_{CR}$ | not specified | 20 |
| $n_e$ (cm$^{-3}$) | 0.03 to 0.32 | |
| C/H | < -3.61 | -4.16 |
| O/H | -3.22 | -3.15 |
| S/H | | -5.0 |
| Si/H | | -5.5 |
| Ca/H | -8.75 | -8.6 |
| K/H | | -8.0 |
| Fe/H | -6.43 | -6.6 |
| Mg/H | -5.35 | -5.3 |
| Na/H† | | -6.5 |
| Ne/H† | | -3.91 |
| Cl/H† | | -7.0 |
| $f$(H$_2$) | 0.44±0.15 | 0.35 |
| Average $\chi^2$ | | 2.4 |

† Assuming ISM abundances from the works of Cowie & Songaila (1986) for the warm and cold phases of the interstellar medium, together with numbers from Table 5 of Savage & Sembach (1996) for the warm and cold phases towards ξ Oph. Our oxygen abundance is taken from Meyer et al. (1998).